\documentclass[twocolumn,prl]{revtex4}
\usepackage{graphicx}
\usepackage{color}
\usepackage{amsmath}

\newcommand{\be}{\begin{equation}}
\newcommand{\ee}{\end{equation}}
\newcommand{\bea}{\begin{eqnarray}}
\newcommand{\eea}{\end{eqnarray}}

\newcommand{\p}{\partial}

\newcommand{\la}{\langle}
\newcommand{\ra}{\rangle}

\newcommand{\lp}{\left(}
\newcommand{\rp}{\right)}

\newcommand{\tr}{{\rm tr\,}}
\newcommand{\bra}[1]{\la #1 |}
\newcommand{\ket}[1]{| #1\ra}

\renewcommand{\vec}[1]{{\bf #1}}

\renewcommand{\phi}{\varphi}
\renewcommand{\epsilon}{\varepsilon}

\newcommand{\addLL}[1]{\textcolor{blue}{#1}}

\begin{document}

\title{Topological Valley Currents in Gapped Dirac Materials}

\author{ Yuri D. Lensky, Justin C. W. Song, Polnop Samutpraphoot, Leonid S. Levitov}

\affiliation{Physics Department, Massachusetts Institute of Technology, 77 Massachusetts Avenue, Cambridge MA02139}

\begin{abstract}
Gapped 2D Dirac materials, in which inversion symmetry is broken by a gap-opening perturbation, feature a unique valley transport regime. The system ground state hosts dissipationless persistent valley currents existing even when topologically protected edge modes are absent or when they are localized due to edge roughness. Topological valley currents in such materials are dominated by 
bulk currents produced by electronic states just beneath the gap rather than by edge modes. 
Dissipationless currents induced by an external bias are characterized by a quantized half-integer valley Hall conductivity.  The under-gap currents dominate magnetization and the charge Hall effect in a light-induced valley-polarized state. 
\end{abstract}

\maketitle
Bloch bands in materials with broken inversion symmetry can feature Berry curvature, an intrinsic physical field  
which dramatically impacts carrier transport\cite{xiao_2010, Nagaosa}.  The key manifestation of Berry curvature is the anomalous Hall effect (AHE), arising in the absence of magnetic field due to topological currents flowing in system bulk transversely to the applied electric field\cite{Sundaram, Haldane}. 
Of high current interest are Dirac materials with several valleys, such as graphene and transition metal dichalcogenide monolayers\cite{xiao_2007, xiao_2012}. Topological currents in these systems have opposite signs in different valleys and, if intervalley scattering is weak, can give rise to long-range charge-neutral valley currents. Such currents have been observed recently in graphene/hBN superlattices\cite{gorbachev_2014}. Alternatively, if valley polarization is induced by light with nonzero helicity, a charge Hall effect is observed\cite{mak_2014}. 

Topological effects are particularly striking in gapped systems where Chern bands 
support topologically  protected  edge modes and quantized transport \cite{kane_2005,bernevig_2006,fu_2007,roy_2007}. 
However, existing Valley Hall materials\cite{xiao_2007, xiao_2012,mak_2014,gorbachev_2014} lie squarely outside this paradigm.  
First, gapless edge states in these materials are not enforced by topology or symmetry and may thus be absent. Second, even when present, these states are not protected against backscattering and localization. Na\"ively, the lack of edge transport would 
lead one to conclude that topological currents cease to exist. 
If true, this would imply that the key manifestations, such as the valley Hall conductivity and orbital magnetization, vanish in the gapped state\cite{xiao_2007}. 

Here we argue that the opposite is true: the absence of conducting edge modes does not present an obstacle since valley currents can be transmitted by 
bulk states beneath the gap. As we will see, rather than being vanishingly small, valley currents 
peak in the gapped state. Further, we will argue that such currents are of a persistent nature, i.e. they represent a ground state property, an integral part of thermodynamic equilibrium. In a valley-polarized state, the under-gap currents dominate magnetization and the charge Hall effect.

\begin{figure}
\centering \includegraphics[scale=0.1]{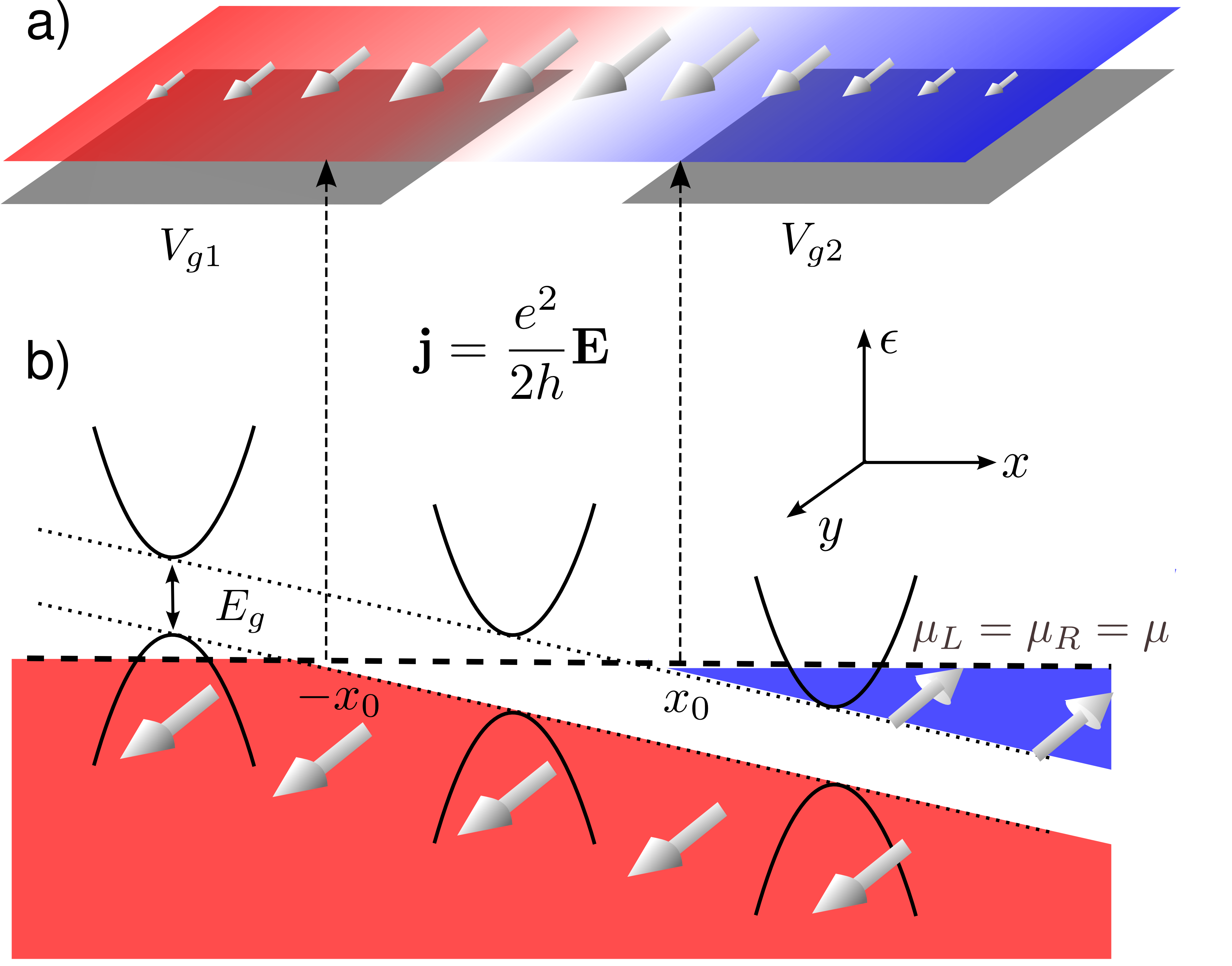}
\caption{Persistent valley currents inside and outside pn junction. The currents 
arise from  side jumps of band carriers just beneath and just above the gap
upon reflection from the gapped region, as illustrated by trajectories in  Fig.\ref{fig2}. The under-gap and over-gap currents 
(red and blue regions) flow in opposite directions and fully cancel deep in the Fermi sea. 
The two contributions are maximally uncompensated inside the region $-x_0<x<x_0$, 
giving a maximum current value of
$j=\frac{e^2}{2h}E$ per valley, where $E$ is the built-in electric field. 
}
\label{fig1}
\vspace{-6mm}
\end{figure}

The effects due to under-gap states, discussed below, should be contrasted with those due to deep-lying states 
which are responsible for field-theoretic anomalies\cite{adler,bell_jackiw}. The anomaly-related currents can lead to interesting transport effects such as the chiral transport in Weyl semimetals\cite{kharzeev,abanin} and in ${}^3$He\cite{volovik}. Importantly,  the deep-lying states in our system obey inversion symmetry and thus do not contribute to valley transport. Indeed, a weak gap-opening perturbation which breaks 
inversion symmetry for states with energies near the Dirac point will have no impact on the deep-lying states. This is quite unlike the anomaly situation where 
symmetry is broken by regularization at the bandwidth scale but remains intact at lower energies. The regime studied here, where valley currents are dominated by 
states just beneath the gap, is unique for systems with a weak inversion-breaking perturbation. A similar behavior is expected in systems such as graphene bilayers in a transverse $E$ field and twisted graphene bilayers.


To gain insight into these delicate issues, we consider a model edge-free gapped system: 
a pn junction in gapped graphene created by a built-in electric field, see Fig.\ref{fig1}. This system features an interesting spatial distribution of valley currents.
As we will see, in contrast to the results of Ref.\cite{xiao_2007}, valley currents reach {\it maximum value} in the gapped pn region $-x_0<x<x_0$. The origin of such (perhaps counterintuitive) behavior is as follows. Valley currents are due to  the states just above and just below the gap and, crucially, are of opposite sign for the two groups of states.  These states are either both depleted of carriers or both filled away from the gapped region, giving 
contributions that nearly cancel. This produces a net current decreasing to zero away from the pn region, see Eq.(\ref{eq:j_spatial}). Within the pn region these contributions are maximally imbalanced, creating a maximum current. Further, the current is quantized to a half-integer value
per valley, $j=\frac{e^2}{2h}E$, where $E$ is the built-in field. 

Our analysis, which is microscopic and explicit, applies equally well to a spatially uniform gapped system under bias (and with no gate-induced built-in field), predicting a quantized Valley Hall effect with $\sigma_{xy}=\frac{e^2}{2h}$ per valley. In terms of the arrangement shown in Fig.\ref{fig1} this corresponds to system sizes $L$ smaller than the gapped region width $2x_0=E_{\rm g}/eE$, i.e. weak bias voltage 
$eV=eEL\ll E_{\rm g}$. 
Since valley currents in this case are transmitted by the under-gap states in the system bulk, they are nondissipative. Below we also discuss 
valley edge currents resulting from the side jumps of the under-gap states upon reflection from system boundary, see Fig.\ref{fig3}. Together with 
bulk currents, such edge currents ensure the valley flow continuity. These currents circulate along the edge, producing orbital magnetization in the system ground state, see Eq.(\ref{eq:magnetization}).

We model carriers in each valley as $2+1$ massive Dirac particle in the presence of a static uniform electric field which defines a pn junction: 
\be\label{eq:H}
H=\lp\begin{array}{cc} \Delta & vp_- \\ vp_+ & -\Delta\end{array}\rp -e E x
,\quad
p_\pm=p_1\pm ip_2
,
\ee
where $p_{1,2}$ denote momentum components $p_{x,y}$. 
The system ground state is a Fermi sea with a density gradient imposed by the $E$ field, n-doped on one side and p-doped on the other side of a gapped region, see Fig.\ref{fig1}. 
Simple as it is, the above Hamiltonian captures all essential elements of interest:  tunneling through the gapped region, AHE in surrounding regions, and their interplay. 

Our approach relies on a mapping onto a fundamental problem in quantum dynamics:  a pair of quantum levels driven through an avoided level crossing. The Landau-Zener (LZ) problem describing these transitions admits an exact solution\cite{LZtransitions}.
The LZ theory provides a very general method that accounts for the AHE transport both outside and inside the gapped pn region, as well as for tunneling through this region. 
Below we discuss the relation between our LZ approach and 
the conventional quasiclassical approach based on the adiabatic theorem and Berry phase\cite{xiao_2010, Nagaosa}. Since the LZ approach is not restricted to the adiabatic limit, it provides a full description of non-adiabatic effects, associated with tunneling through the gapped region in our transport problem. Such effects, which are naturally described in the LZ framework, are not accounted for by the quasiclassical approach.

\begin{figure}
\centering \includegraphics[scale=0.12]{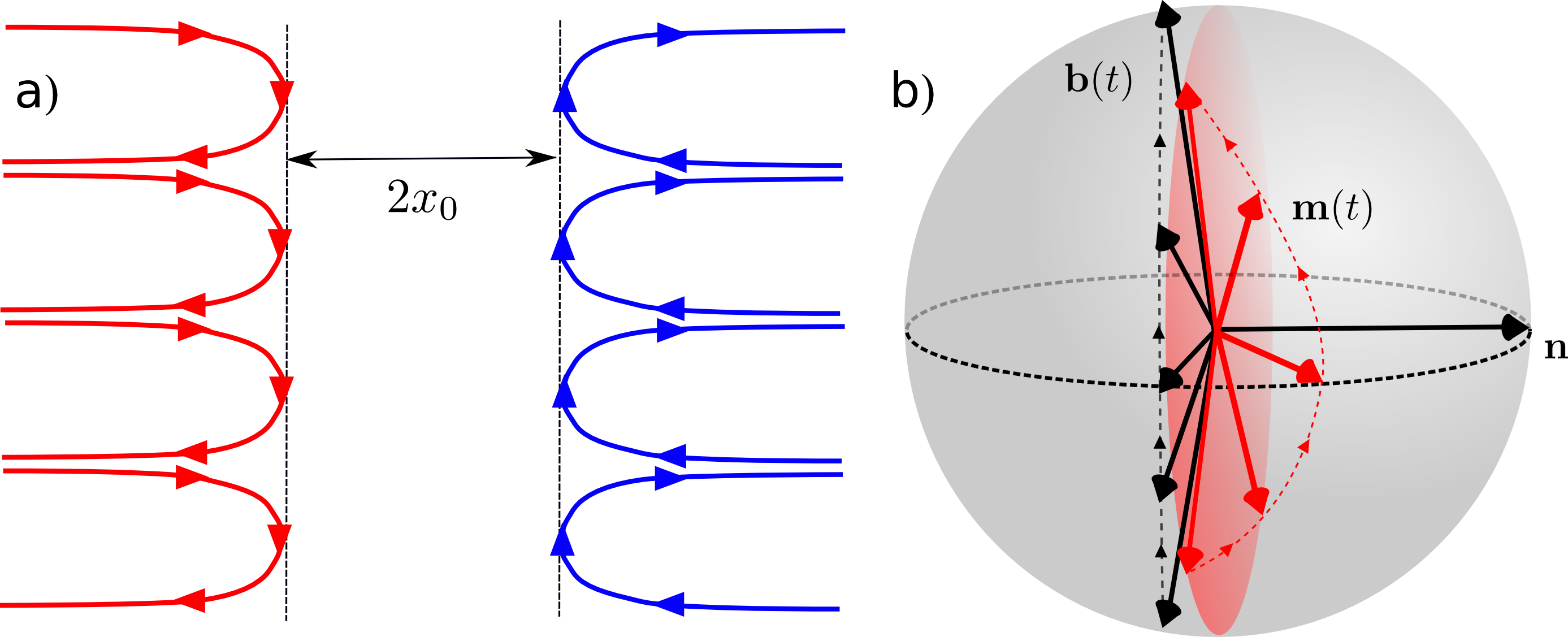} 
\caption{a) The under-gap and over-gap trajectories near the gapped region, Eq.(\ref{eq:y(t)}). Skewed Hall-like motion gives rise to side jumps.  Shown are normally incident trajectories (red for electrons, blue for holes). 
The opposite-flowing under-gap and over-gap currents partially cancel when summed over all filled states, producing net currents flowing in the same direction in the p and n regions, with the maximum current attained in the middle region $-x_0<x<x_0$, see Eq.(\ref{eq:j_spatial}) and Fig.\ref{fig1}. b) Spin-1/2 interpretation of side jumps. Magnetization $\vec m(t)$ evolves adiabatically
in a slowly varying field $\vec b(t)$ that sweeps a plane perpendicular to $\vec n$, see Eq.(\ref{eq:bloch_eqn}). Magnetization tracks the field but lags slightly behind, rotating out of the plane and acquiring a component parallel to $\vec n$, see Eq.(\ref{eq:b_eff}). So does the velocity vector which is aligned with $\vec m(t)$.
}
\label{fig2}
\vspace{-6mm}
\end{figure}

Mapping of Eq.(\ref{eq:H}) onto the LZ problem is  accomplished in two steps. We first note that in the momentum representation $\epsilon\psi=H\psi$ is a first-order differential equation, since the only term containing a derivative is $-eEx=eEi\hbar \p_{p_1}$. 
We can thus rewrite our equation as a time-dependent Schroedinger equation for 
a $2\times 2$ Hamiltonian, 
with $t=p_1/eE$ playing the role of time:
\be\label{eq:H_spin}
i\hbar\p_t \psi(t)=\tilde H(t)\psi(t) 
,\ \ \
\tilde H(t)=
\beta t\sigma_1+vp_2\sigma_2+\Delta\sigma_3
,
\ee
where  we set $\epsilon=0$ without loss of generality and defined $\beta=veE$. Next, by interchanging spin components via $\sigma_1\leftrightarrow\sigma_3$, $\sigma_2\to -\sigma_2$ we bring $\tilde H$ to the canonical LZ form
\be\label{eq:H(t)}
\tilde H(t)=\lp\begin{array}{cc}\beta t &\Delta_p \\ \Delta_p^* & -\beta t\end{array}\rp
,\quad
\Delta_p=\Delta +ivp
,
\ee
where from now on we use $p$ instead of $p_2$ for brevity. 

Time evolution in Eq.(\ref{eq:H_spin}) defines a unitary S-matrix which takes its simplest form in the adiabatic basis of instantaneous eigenstates of $\tilde H(t)$. These states correspond to a particle moving in a classically allowed region, p or n, without tunneling through the gapped region. Tunneling is thus described by the LZ  transitions between different adiabatic states.
Written in the adiabatic basis, the S-matrix is  of the form
\be\label{eq:S-matrix}
S=\lp\begin{array}{cc} \sqrt{q} & -\sqrt{1-q}e^{i\phi} \\ \sqrt{1-q}e^{-i\phi} & \sqrt{q}\end{array}\rp
,\quad
q=e^{-2\pi\delta}
,
\ee
where 
$\delta=|\Delta_p|^2/2\beta\hbar$. Here the phase $\phi$ is given by\cite{kayanuma94}
\be\label{eq:phi}
\phi=\pi/4+\arg \Gamma(1-i\delta)+\delta(\ln\delta-1) + \arg \Delta_p
\ee
with $\Gamma(z)$ the Gamma function.  The non-adiabatic and adiabatic LZ transitions, 
taking place with the  probabilities $q$ and $1-q$, correspond to particle transmission through the gapped region and reflection from it. The evolution is adiabatic at small $\beta$, with the system tracking one of the instantaneous eigenstates of $\tilde H(t)$ and  non-adiabatic transitions exponentially suppressed, $q\to 0$.

The S-matrix exhibits features characteristic for the skewed particle motion taking place in the AHE regime. In particular, it predicts side jumps --- particle transverse displacement 
induced by its proximity to the pn region. We evaluate the $y$ displacement as $\la \delta y\ra=\bra{\psi} i\hbar\p_p\ket{\psi}$ 
with the expectation value taken over the left- and right-incident states, 
$\ket{L}=S\lp\begin{array}{c} 1\\0 \end{array}\rp$, $\ket{R}=S\lp\begin{array}{c} 0\\1 \end{array}\rp$. We find
\be\label{eq:delta_y}
\la \delta y\ra_{L,R}=\pm\p\phi/\p p=\pm\ell (1-q)
,\quad
\ell=\hbar v\Delta/|\Delta_p|^2
,
\ee
where only the last term of the phase in Eq.(\ref{eq:phi}) gives a contribution even in $p$ contributing to the net valley current. Interestingly, the result in Eq.(\ref{eq:delta_y}) only depends on $1-q$ that corresponds to reflection, 
 indicating that side jumps occur only at reflection from the gapped region but not at transmission through it. 
The side jump direction reverses upon reversing the sign of $\Delta$, giving values of opposite sign for valleys $K$ and $K'$ as expected for Valley Hall transport. 


Encouraged by these observations, here we construct individual one-particle quantum states exhibiting side jumps. 
Since Dirac particle velocity is expressed through its spin, $\vec v=\frac1{i\hbar}[\vec x,H]=v(\sigma_1,\sigma_2)$, it will be convenient to represent LZ dynamics as spin $1/2$ evolution. The latter is described by 
the Bloch equation for magnetization vector $\vec m (t)=\bra{\psi(t)}\vec s\ket{\psi(t)}$, $s_i=\frac{\hbar}2\sigma_i$, 
\be\label{eq:bloch_eqn}
\p_t\vec m =\vec b(t)\times\vec m
,\quad
\vec b(t)=\frac2{\hbar}(\Delta,-vp,\beta t)
\ee
where the magnetic field $\vec b(t)$ orientation changes from $-z$ to $+z$ over $-\infty<t<\infty$. 

We focus on the weak field regime $eE\ll\Delta/\ell=\Delta^2/\hbar v$. In the LZ formulation (\ref{eq:H(t)}) this corresponds to spin $1/2$ evolving in a slowly changing magnetic field $\vec b(t)$ which rotates in the plane perpendicular to the vector 
\be\label{eq:n_alpha}
\vec n=(\sin\alpha,\cos\alpha,0)
,\quad
\tan\alpha=vp/\Delta
.
\ee
Crucially, the adiabatic spin evolution in a rotating field $\vec b(t)$ can generate a component of $\vec m$ (and thus of the velocity) transverse to the rotation plane and thus {\it pointing along } $\vec n$. This happens because when the field rotates 
in the plane perpendicular to $\vec n$ the spin tries to follow it but is left slightly behind. Then, as a result of Bloch precession, the spin rotates out of the plane swept by $\vec b(t)$, see Fig.\ref{fig2}. This component is proportional to rotation speed, i.e. is not exponentially small in the adiabatic limit. 

Such a behavior, while  somewhat counterintuitive, can be understood as follows. We usually think of a spin precessing in a strong but slowly changing magnetic field is being ``slaved to the field''. This is basically correct, however the spin excursions away from the field direction can be nonexponential due to the Berry curvature effects. This is precisely the case in our problem. 

It is convenient to use a (nonuniformly) rotating frame in which the field $\vec b(t)$ has a frozen orientation. 
We write $\ket{\psi(t)}=U(t)\ket{\psi'(t)}$ 
with the unitary transformation $U(t)$ chosen so that the field $\vec b'(t)$  defined by 
$
U^{-1}(t)\lp \vec b(t)\cdot \vec s \rp U(t)=\vec b'(t)\cdot \vec s
$
is directed along a fixed axis. For the Hamiltonian in Eq.(\ref{eq:H(t)}) the operator $U(t)$ with this property can be defined as a spin rotation 
\be
U(t)=e^{ \frac{i}{\hbar}\theta(t)\vec n\cdot \vec s}
,\quad
\tan\theta(t) = \beta t/|\Delta_p|
,
\ee
where $\theta(t)$ is the angle between vectors $\vec b(t)$ and $\vec b(0)=\frac2{\hbar}(\Delta, -vp,0)$.
In the rotated frame our equations read
\be
i\hbar\p_t\ket{\psi'(t)}=\lp \vec b'(t)\cdot\vec s-i\hbar U^{-1}(t) \dot U(t) \rp \ket{\psi'(t)}
.
\ee
The last term equals
$
-i\hbar U^{-1}(t) \dot U(t)=\frac{\p \theta(t)}{\p t}\vec n\cdot \vec s
$
giving a spin Hamiltonian with an effective field $\vec b'(t)+\frac{\p \theta(t)}{\p t}\vec n$. 

So far our analysis has been completely general, now we specialize to 
an adiabatic evolution in which the spin orientation tracks 
the field. In this case, when viewed in our rotated frame, $\vec m(t)$ remains aligned 
with the vector $\vec b'(t)+\frac{\p \theta(t)}{\p t}\vec n$ at all times. 
Transforming back to the lab frame, we conclude that $\vec m(t)$ tracks the field 
\be\label{eq:b_eff}
\tilde{\vec b}(t)=\vec b(t)+\frac{\p \theta(t)}{\p t}\vec n
\ee 
which, because of the last term, has an additional $y$ component. 
Finally, since the velocity operator $\vec v=v(\sigma_1,\sigma_2)$ expectation value is aligned with $\vec m$, the velocity  components are easily evaluated as $\vec v_{x,y}=v\tilde{\vec b}_{x,y}/|\tilde{\vec b}|$ giving
\be\label{eq:v(t)}
v_x(t)=\frac{v\beta t}{\epsilon(t)}
,\quad
v_y(t)=\frac{v^2 p}{\epsilon(t)}+\frac{v\Delta \beta}{2\epsilon^3(t)}
,
\ee
where $\epsilon(t)=\pm \sqrt{\beta^2 t^2+|\Delta_p|^2}$ with the plus/minus sign describing p and n states. Here we normalized $\tilde{\vec b}(t)$ approximating $|\tilde{\vec b}(t)|\approx |\vec b(t)|$. Trajectories are readily obtained  by integrating velocity, giving
\be\label{eq:y(t)}
x(t)=\frac{v}{\beta}\epsilon(t)
,\quad
y(t)=\frac{v^2 p \ln\frac{\epsilon(t)+\beta t}{|\Delta_p|}}{|\Delta_p|}+\frac{v\Delta\beta t}{2|\Delta_p|^2\epsilon(t)}
\ee
(here we have suppressed integration constants). 
The last term in Eqs.(\ref{eq:v(t)}),(\ref{eq:y(t)}) 
originates from Berry curvature, giving rise to side jumps, see Fig.\ref{fig2}. 
 The net side jump value is 
$\delta y=\int_{-\infty}^\infty v_y(t)dt=v\Delta/|\Delta_p|^2$, which matches the result found above for $p=0$.

These results are in accord with the classical equations of motion augmented with the anomalous velocity term 
describing the nonclassical Berry's ``Lorentz force:"\cite{xiao_2010, Nagaosa}
\[
\dot {\vec p}=e\vec E,\quad
\dot{\vec x}=\nabla_{\vec p} \epsilon_\pm(\vec p)+\Omega(\vec p)\times\dot {\vec p}
,\quad
\Omega(\vec p)=\frac{v^2\Delta }{2\epsilon^3_\pm(\vec p)}
,
\]
where $\epsilon_\pm(\vec p)=\pm (v^2\vec p^2+\Delta^2)^{1/2}$ is 
particle dispersion. 
Current density, found by summing the velocity contributions of all states in the Fermi sea, is
\be\label{eq:j_spatial}
j(x)=\left\{\begin{array}{ll} j_0, & |x|<x_0\\ j_0 x_0/|x|, & |x|>x_0\end{array}\right.
,\quad 
j_0=\frac{e^2}{2h}E
\ee
per valley. The current peaks in the gapped region, falling off inversely with distance outside this region, as shown in Fig.\ref{fig1}a. An identical result is obtained by integrating the velocity in Eq.(\ref{eq:v(t)}) 
over allowed values of $p$. As discussed above, this behavior originates microscopically from the contribution of the under-gap trajectories side jumps dominating in the gapped region, however being partially canceled by the over-gap trajectories contribution outside this region. 
Interestingly, the linear dependence $j_0$ vs. $E$ translates into a universal, $E$-independent net current flowing through the gapped region, $I_{|x|<x_0}=\frac{e\Delta}{\hbar}$. This behavior is linked to the side-jump values being independent of $E$ at weak field.

Dissipationless currents in a spatially uniform gapped system can also be created by a voltage bias. The $E$-independent and small side-jump values noted above allow us to treat a gapped system under a weak bias (and no gate-induced built-in fields) using the above model, as long as $eE\ll\Delta/L$ where $L$ is system size. Our analysis then predicts a universal valley Hall conductivity 
$\sigma_{xy}=\frac{e^2}{2h}$ per valley. Since valley currents in this case are transmitted solely by 
under-gap states in the system bulk, they are nondissipative. 

Another interesting phenomenon is {\it persistent edge currents} in a spatially uniform unbiased gapped system. These currents arise via a similar mechanism, 
due to side jumps of the under-gap states scattered off system edges, see Fig.\ref{fig3}. 
These currents circulate along the edge, producing orbital magnetization in the system ground state. The $K$ and $K'$ valley contributions are of opposite sign, giving zero net  magnetization in thermodynamic equilibrium. Finite total magnetization can be created by using light of a particular helicity to polarize valleys 
(as in the Valley Hall effect measurements\cite{xiao_2007,mak_2014}), 
\be
\label{eq:magnetization}
M\!\!=\!\int \!\! d^2 r \frac1{2c} {\vec r}\!\times \vec j(\vec r)
\approx A \frac{\gamma e\Delta}{\hbar c}
\sum_{\vec p,i,\pm} \Omega(p)n_{i,F}(\epsilon_\pm(p))
,
\ee
where $A$ is system area, $n_{i,F}$ are the Fermi functions with $i$ labeling valleys, and $\gamma\sim 1$ a numerical constant accounting for edge current suppression due to intervalley scattering induced by edge roughness. This estimate was obtained by setting the typical side-jump value equal to that found for the pn region. The dependence on the Fermi level arises from summing the contributions of all filled states. Magnetization attains maximum value when the Fermi level lies inside the gap, see Fig.\ref{fig3}. 
In 2D systems, magnetization can be measured with 
torque magnetometry techniques, allowing access to 
values as low as $0.1\,\mu_B$/2D u.c. \cite{ashoori_magnetization}.

\begin{figure}
\centering \includegraphics[scale=0.22]{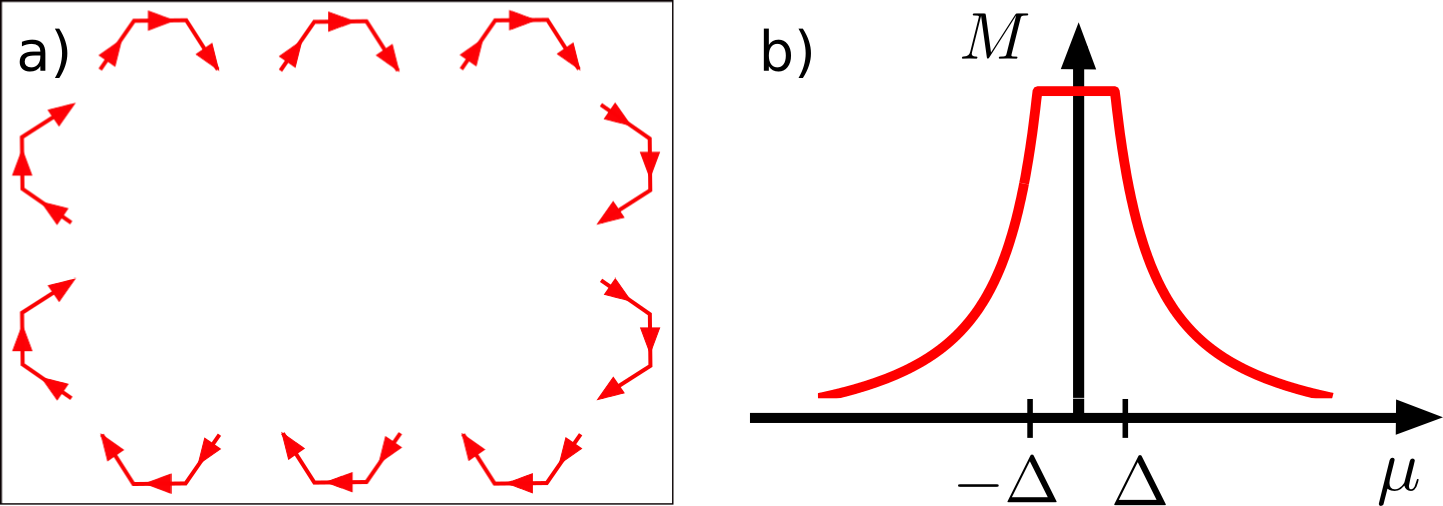} 
\caption{a) Persistent valley currents in a spatially uniform gapped system with the Fermi level inside the gap. Currents arise due to side jumps of the under-gap trajectories bouncing off the system boundary. 
Persistent currents circulate along the edge, giving rise to a constant magnetization per valley, Eq.(\ref{eq:magnetization}).
b) Orbital magnetization, Eq.(\ref{eq:magnetization}), as a function of chemical potential. Magnetization reaches maximum value for the Fermi level inside the gap and decreases to zero at large detuning as a result of compensation from over-gap and under-gap contributions.
}
\label{fig3}
\vspace{-6mm}
\end{figure}

In conclusion, topological valley currents in gapped materials are not transmitted by edge modes, which are not protected by topology or symmetry, but rather by 
under-gap bulk states. We demonstrate that the absence of conducting edge modes does not present an obstacle since the under-gap currents can give rise to dissipationless transport in the gapped state. The under-gap currents generate persistent (magnetization) currents in the thermodynamic ground state, flowing in the system bulk and along boundaries. We predict that the key manifestations and observables, such as the 
Valley Hall conductivity and orbital magnetization in valley-polarized systems, reach maximum value in the gapped state.  The requirements for observing dissipationless valley transport can be met under realistic conditions. 

%
%
%
%
%
%

We thank R. C. Ashoori, A. K. Geim, and P. L. McEuen for useful discussions. This work was supported, in part, by STC Center for Integrated Quantum Materials, NSF grant DMR-1231319.

\end{document}